\documentclass[conference]{IEEEtran}
\ifCLASSINFOpdf
\else
\fi

\usepackage{tikz}
\usetikzlibrary{plotmarks,shapes,patterns,decorations.pathreplacing,backgrounds,calc,arrows,arrows.meta,spy,matrix,shadows,trees,positioning,fit}

\usepackage{todonotes}
\usepackage{lipsum}  
\usepackage{xcolor} 

\usepackage{float} 
\usepackage{caption} 
\usepackage[draft]{hyperref}
\usepackage{glossaries}
\usepackage{xcolor}
\usepackage{subfig}
\usepackage{balance}
\hypersetup{
    colorlinks,
    linkcolor={black},
    citecolor={black},
    urlcolor={blue}
}

\newacronym{leo}{LEO}{low Earth orbit}
\newacronym{rtt}{RTT}{Round-Trip Time}
\newacronym{ntn}{NTN}{Non-Terrestrial Network}
\newacronym{tcp}{TCP}{Transmission Control Protocol}
\newacronym{ccdf}{CCDF}{complementary CDF}
\newacronym{nce}{NCE}{Non-Congestion Events}
\newacronym{ncl}{NCL}{Non-Congestion Loss}
\newacronym{rr}{RR}{Reordering-Robust}
\newacronym{door}{DOOR}{Detecting Out of Order and Response}
\newacronym{fso}{FSO}{Free-Space Optical}
\newacronym{icmp}{ICMP}{Internet Control Message Protocol}
\newacronym{bbr}{BBR}{Bottleneck Bandwidth and Round-trip propagation time}
\newacronym{cdf}{CDF}{Cumulative Distribution Function}
\newacronym{dash}{DASH}{Dynamic Adaptive Streaming over HTTP}
\newacronym{isl}{ISL}{inter-satellite link}
\newacronym{e2e}{E2E}{end-to-end}
\newacronym{gs}{GS}{Ground Station}
\newacronym{qos}{QoS}{Quality of Service}
\glsdisablehyper
\begin{document}
%
\title{End-to-End Delivery in LEO Mega-constellations and the Reordering Problem}


\author{\IEEEauthorblockN{Rasmus Sibbern Frederiksen\IEEEauthorrefmark{1}, Thomas Gundgaard Mulvad\IEEEauthorrefmark{1}, Israel Leyva-Mayorga\IEEEauthorrefmark{1}, \\Tatiana Kozlova Madsen\IEEEauthorrefmark{1}, and Federico Chiariotti\IEEEauthorrefmark{2}}
\IEEEauthorblockA{\IEEEauthorrefmark{1} Department of Electronic Systems, Fredrik Bajers Vej 7C, 9220 Aalborg, Aalborg University, Denmark\\
\IEEEauthorrefmark{2} Department of Information Engineering, via Giovanni Gradenigo 6B, 35131 Padua, University of Padova, Italy \\
Emails: \{rsfr, ilm, tatiana\}@es.aau.dk, tmulva19@student.aau.dk, chiariot@dei.unipd.it
} 
}


%


\maketitle
\begin{abstract}
\Gls{leo} satellite mega-constellations with hundreds or thousands of satellites and \glspl{isl} have the potential to provide global end-to-end connectivity. Furthermore, if the physical distance between source and destination is sufficiently long, end-to-end routing over the \gls{leo} constellation can provide lower latency when  compared to the terrestrial infrastructure due to the faster propagation of electromagnetic waves in space than in optic fiber. However, the frequent route changes due to the movement of the satellites result in the out-of-order delivery of packets, causing sudden changes to the \gls{rtt} that can be misinterpreted as congestion by congestion control algorithms. In this paper, the performance of three widely used congestion control algorithms, Cubic, Reno, and BBR, is evaluated in an emulated \gls{leo} satellite constellation with \gls{fso} \glspl{isl}. Furthermore, we perform a sensitivity analysis for Cubic by changing the satellite constellation parameters, length of the routes, and the positions of the source and destination to identify problematic routing scenarios. The results show that route changes can have profound transient effects on the goodput of the connection, posing problems for typical broadband applications.


\end{abstract}

\glsresetall

\section{Introduction}
\label{sec:intro}

Compared to terrestrial base stations, satellites achieve a much wider coverage due to their high altitude of deployment. Additionally, the availability of satellite infrastructure does not depend on the accessibility of the terrain nor on the availability of existing road and power infrastructure, making Internet access feasible in regions with a shortage of cables. Despite these advantages, in the previous decades, satellites were mainly utilized for TV broadcasting and telephony, whereas satellite Internet access was restricted to otherwise disconnected areas, such as the north of Greenland~\cite{SimonsenAbildgaard2022, Chotikapong2001}, as it was seen as a sub-par alternative to conventional Internet in terms of performance and cost.  However, there is a renewed enthusiasm for satellite communications driven by developments in \gls{ntn} wireless communications technology and standards~\cite{3GPPTR38.821}. These include the Release 17 of 5G New Radio (NR)~\cite{3GPPTS38.300}, as well as space-focused technologies leading to a sharp drop in the cost of rocket launches. Consequently, satellite communications will play a key role in future communication systems~\cite{3GPPTS38.300, Liu2021}. 

\begin{figure}[t]
    \centering
    \includegraphics[width=0.85\linewidth]{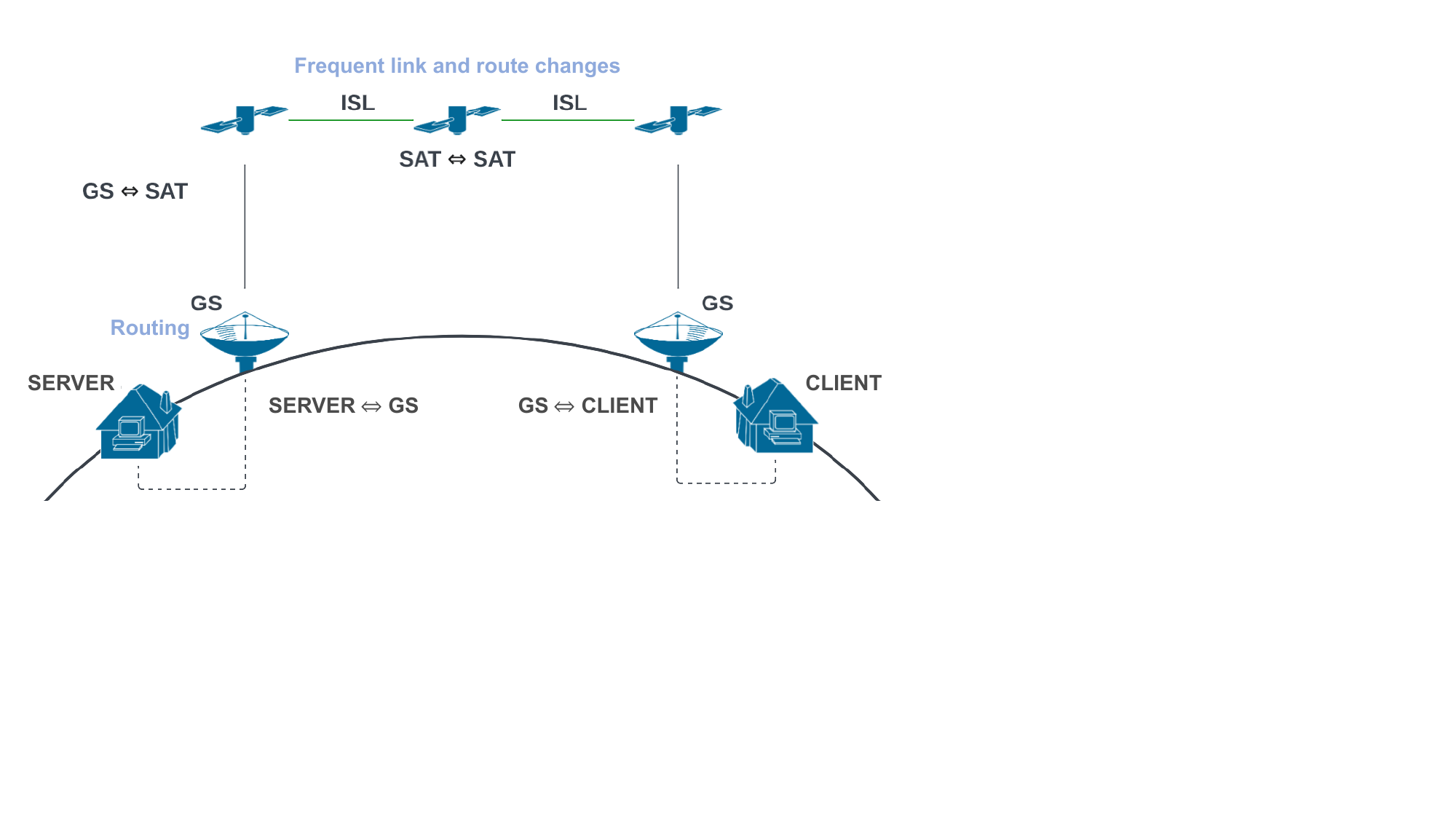}\vspace{-1em}
    \caption{Considered use case of remote Internet access, connecting remote \glspl{gs} through a \gls{leo} satellite constellation with \glspl{isl}.}\vspace{-0.4cm}
    \label{fig:satellite_internet}
\end{figure}

One of the major drivers for this so-called New Space era is the fact that low latency and reliable Internet access from space can be provided by deploying a constellation of satellites in \gls{leo}, between $500$ and $2000$ km above the Earth's surface. Consequently, there is an unprecedented opportunity to commercialize high-quality satellite Internet services through \gls{leo} constellations, for example, by serving as a backhaul network in remote or disaster-struck areas as shown in Fig.~\ref{fig:satellite_internet}, or to offload the ground infrastructure in dense urban areas~\cite{Soret2021}. 

In the past few years, several companies have heavily invested in their so-called \gls{leo} mega-constellations, including SpaceX's  Starlink constellation of nearly $12000$ planned satellites, and OneWeb. These will soon be joined by several other companies, such as Amazon with Project Kuiper and Telesat. Previous studies have concluded that \gls{leo} mega-constellations with \gls{fso} \glspl{isl} can be used for low-latency \gls{e2e} connectivity across large distances, as propagation of electromagnetic waves is faster in space than in optic fiber by a factor determined by its refraction index~\cite{Handley2018}. Furthermore, \gls{fso} technologies have evolved to provide high data rate communication at the large distances that occur in \gls{leo} constellations~\cite{Kaushal2017} and it has been observed that \gls{fso} \glspl{isl} may greatly enhance the transport capacity of \gls{leo} mega-constellations~\cite{DelPortillo2019}. However, these conclusions have been drawn based on the distances traversed by the links and without considering that the evolution in congestion control algorithms for \gls{ntn} has not been so rapid and these might be the true bottleneck of the \gls{leo} systems.


A main challenge to achieve global Internet access through \gls{leo} constellations is the mobility of the infrastructure. \Gls{leo} satellites must orbit the Earth at around $7.5$\,km/s to maintain their orbit and, hence, the topology of \gls{leo} constellations changes constantly. In L1 and L2, this mobility creates the need for frequent link adaptation, antenna pointing, and link establishment procedures~\cite{Leyva-Mayorga2021}. In L3, this mobility creates the need for frequent route adaptations, which end up affecting the performance of L4 protocols such as the widely used \gls{tcp}, which is an important component of the current terrestrial Internet architecture. \Gls{tcp} is a connection-oriented protocol in the Internet Protocol suite (TCP/IP), which is responsible for reliable \gls{e2e} delivery of the data following a chronological order. \gls{tcp} contains several mechanisms to maintain and ensure reliable data delivery in packet-based communication, solving problems such as congestion and out-of-order reception. There are many variants of \gls{tcp} congestion control algorithms, including the classic Reno, Vegas, and Cubic, and targeted improvements have been done to support satellite communications~\cite{RFC2488}. 

However, \gls{tcp} was designed for fixed terrestrial networks and the improvements for satellite communications did not consider the large scale of modern \gls{leo} satellite mega-constellations, which \gls{tcp} might experience severe performance degradation. In particular, various features of \gls{leo} constellations, including the variable \gls{rtt}, pose problems to \gls{tcp}~\cite{Chotikapong2001, Wood2001, Giambene2018}. Nevertheless, most of the works on performance analysis of \gls{tcp} are based on the long-term average response time and the duration of the transmissions~\cite{Chotikapong2001}, average goodput~\cite{Giambene2018}, or the progress of file transfers~\cite{Wood2001}, which do not provide insights in the short-term stability of the connection. Currently, only a few constellations have implemented \glspl{isl} and the L4 mechanisms that are implemented in them, including Starlink, are not publicly available. Because of this, it not possible to ascertain the L4 mechanisms that are currently used in these networks nor to draw insightful conclusions beyond the result of the experiments. For instance, a few works have investigated the short-term stability of the connection in Starlink and have found that the throughput is affected every $15$\,ms. This problem is attributed to periodic reconfiguration processes but, due to the lack of information on the implemented mechanisms, the root cause is yet to be identified\,\cite{Mohan2023}. Nevertheless, the early stage of adoption of \glspl{isl} raises the importance of investigating the L4 mechanisms and their variants that are able to achieve the best long- and short-term performance in satellite mega-constellations.  

In this paper, we conduct a detailed performance evaluation of \gls{tcp} with three congestion control algorithms, Reno, Cubic, and Google's novel \gls{bbr} algorithm~\cite{cardwell2016bbr}, through emulation. The emulator was developed for this purpose using netem and has been made freely available online\,\footnote{\url{https://github.com/rasmussibbern88/satellite_tcp_emulator}.}. We focus on remote Internet access scenarios where data is routed \gls{e2e} through a \gls{leo} constellation with the original OneWeb geometry. Furthermore, we perform a sensitivity analysis on the constellation geometry and the distance and position between the source and destination. We observe that the transient effects after route changes can last between a few hundred milliseconds and a couple of seconds, where goodput is severely affected. Therefore, these transient effects may lead to \gls{qos} degradation in typical multimedia services.

To the best of our knowledge, this is the first extensive analysis of the behavior of transport layer congestion control mechanisms over \gls{leo} connections, and raises significant concerns that should be addressed by the research and industrial communities. The performance study also highlights the difference between existing mechanisms, which exhibit different performance and failure modes. The main contribution of this work is the analysis of the effects of route changes on modern congestion control algorithms, which exhibit different patterns in terms of delay, \gls{rtt}, and number of retransmitted packets.

\begin{figure}[t]
\centering
\input{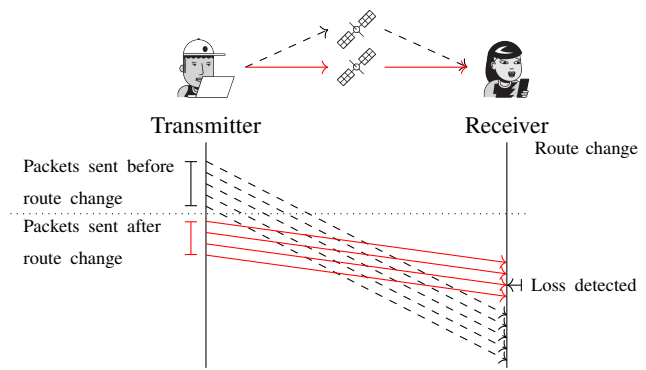}\vspace{-1em}
\caption{Packet-level diagram of the reordering problem.}\vspace{-0.4cm}
\label{fig:reordering}
\end{figure}

\section{The Reordering Problem}
\label{sec:tcp}

One of the critical design choices in \gls{tcp} is the first-in, first-out assumption~\cite{polese2019survey}. In most network scenarios, if packet $A$ is sent before packet $B$, there are three possible outcomes:
\begin{enumerate}
\item The acknowledgment for packet $A$ reaches the sender before the ACK for packet $B$, or the two are acknowledged together;
\item Packet $A$ is lost, and the receiver either sends a duplicate acknowledgment (DUPACK) or a selective acknowledgment (SACK) for packet $B$;
\item Both packets are lost, and no feedback is transmitted.
\end{enumerate}
\emph{Packet reordering} occurs when packet $B$ reaches the receiver before packet $A$, which is not tolerated by the \gls{tcp} design. Consequently, modern \gls{tcp} error recovery mechanisms consider DUPACKs (which are triggered any time packets are received out of order) as a signal of packet loss. The threshold to trigger a retransmission in most \gls{tcp} versions is 3 consecutive DUPACKs, to allow for minor reordering, which is generally enough for static network infrastructures, such as cellular networks~\cite{schulte2017ll}. However, \emph{ad hoc} networks might break this assumption in a much more significant way~\cite{leung2007overview}: in these radio networks with fast mobility, route changes are often frequent, and the changes in the \gls{rtt} can cause significant reordering. Fig.~\ref{fig:reordering} shows a simple diagram of a reordering event: a route change due to the satellites' movement results in a shorter \gls{rtt}, and newer packets arrive before older ones on the previous route: this is perceived as a packet loss event, triggering an expensive retransmission and loss recovery procedure. High-performance networks with load-balancing routing can also cause reordering due to frequent route changes~\cite{carpa2017evaluating}, either because the new route has a lower propagation time, due to a more direct path from the source to the destination, or because it has a lower load, and thus shorter buffers and a lower waiting times for packets. Whatever the cause of the reordering event, its effects on \gls{tcp}'s operation are massive, as this is not a circumstance foreseen by the original protocol design.

 The Eifel algorithm~\cite{ludwig2003eifel} allows \gls{tcp} senders to detect spurious loss recovery conditions, i.e., misinterpreted situations in which the sender retransmits packets unnecessarily, but massive reorderings are assumed to be a rare occurrence. There are several other mechanisms to mitigate this issue in the \gls{tcp} literature, mostly applying to ad hoc networks and based on two alternative principles: delay or reversion. Delay-based techniques solve the reordering problem by responding less aggressively to possible congestion signals, i.e., trying to detect whether a series of DUPACKs is due to congestion or to some other cause, and reacting only in the former case. This mitigates the consequences of reordering events, but also slows down the response to real congestion, causing the sender to behave more aggressively and cause longer delays and more buffer overflows. This family of schemes includes \gls{tcp} \gls{ncl}~\cite{lai2009tcp} and \gls{nce}~\cite{sreekumari2011tcp}, which aim at detecting anomalies (such as multiple consecutive DUPACKs, or long timeouts) that are not caused by congestion losses, reducing the congestion window only when necessary (i.e., in real congestion events). This includes both lower-layer losses and reorderings, and the mechanisms are based on the earlier \gls{tcp} \gls{door}~\cite{wang2002improving}, which turned off congestion control entirely during reordering events. Finally, \gls{rr}~\cite{zhang2003rr} attempts to control the number of DUPACKs required to trigger loss recovery mode, considering the impact and frequency of reordering events. Reversion-based techniques, on the other hand, try to detect the reordering after the fact, reverting to the previous congestion window after the fact. \gls{door}, \gls{ncl}, and techniques based on the Eifel algorithm also include reversion mechanisms, enhancing their adaptability.

 \begin{figure}[t]
    \centering
    \includegraphics[width=0.95\columnwidth,clip,trim={0 0 0 0.75cm}]{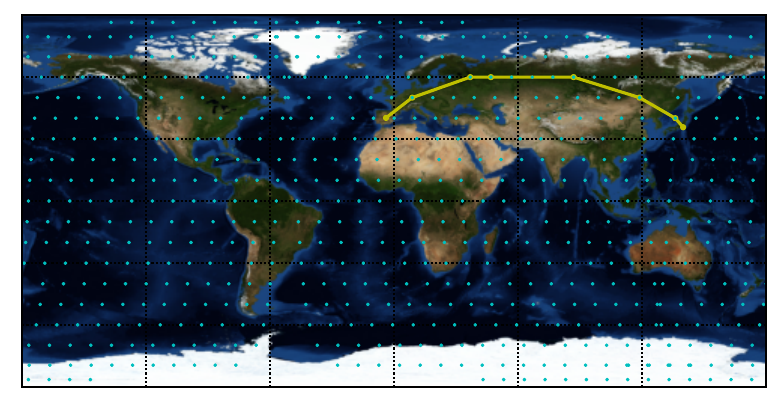}\vspace{-0.3em}
    \caption{Baseline scenario: Shortest-path routing through OneWeb's \gls{leo} constellation between Madrid and Tokyo.}\vspace{-1em}
    \label{fig:routing}
\end{figure}

\begin{table}[tbp]
    \centering\footnotesize
\caption{Parameter settings for performance evaluation.}
\renewcommand{\arraystretch}{1.2}
    \begin{tabular}{@{}ll@{}}
    \hline
    \textbf{Parameter} & \textbf{Value } \\
    \hline
        Altitude of deployment& $1200$\,km \\
        Number of orbital planes & $\{9,18\}$\\
        Number of satellites per orbital plane & $36$\\
        Orbit inclination & $86.4^\circ$\\
        L3 update interval & 30 seconds \\
        L2 update interval & 15 seconds \\
        Routing metric & Shortest distance \\
        FSO link range & $3000$\, km \\
        FSO link data rate & $100$\,Mbps\\
        TCP variants & Cubic, Reno, BBR\\
        Emulation time & $1000$\,seconds\\
        Goodput sampling frequency & $20$\,Hz\\
    \hline
    \end{tabular}\vspace{-0cm}
    \label{tab:emulation_parameters}
\end{table}

 \gls{leo} networks present a new challenge for transport layer research: on one hand, reordering events are frequent because of the fast-varying network topology, while on the other, these shifts are often \emph{predictable}. Indeed, the fast orbital velocity of \gls{leo} satellites results in up to a few route changes per minute. The high capacity of \gls{fso} links between satellites worsens the problem by increasing the number of packets in flight, and consequently the number of packets affected by reordering if the new route is faster. A massive reordering involving hundreds of packets cannot be mitigated by adapting the threshold to trigger a retransmission, as such a high threshold would completely break congestion control and loss detection. However, satellite mobility can be computed days in advance, and in-network techniques to mitigate the problem or alert the endpoints to route changes through the \gls{icmp} can be deployed effectively. The objective of this work is to gauge the extent of the problem in a real \gls{leo} constellation under ideal conditions, so as to raise awareness of the problem in the satellite network community, which has mostly concentrated on the lower layers in the past few years, and provide a solid basis for future network adaptations and solutions to the issue.

\begin{figure}[t]
    \centering
    \includegraphics[width=0.75\columnwidth]{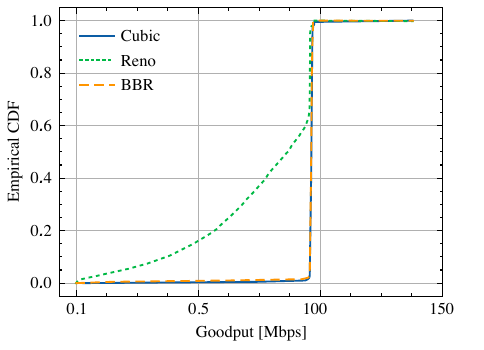}
    \caption{Empirical CDF of the goodput (x-axis in non-linear scale).}
    \label{fig:gp_cdf}\vspace{-0.4cm}
\end{figure}

\begin{figure}[t]
    \centering
    \includegraphics[width=0.75\columnwidth]{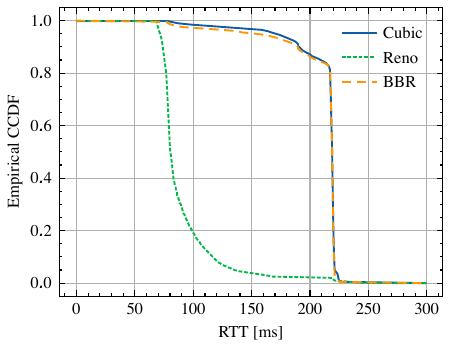}
    \caption{Empirical \gls{ccdf} of the \gls{rtt}.}\vspace{-0.4cm}
    \label{fig:rtt_ccdf}
\end{figure}

\begin{figure*}
    \centering
    \subfloat[Reno\label{sf:long_S_reno}]{\includegraphics[width=0.3\linewidth]{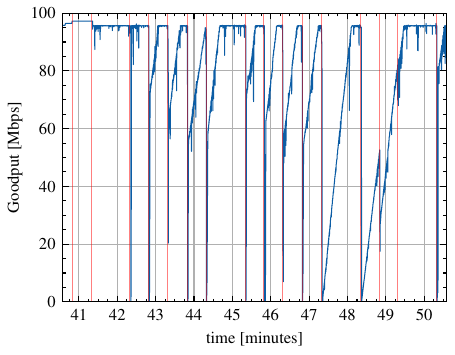}}\hfil
    \subfloat[Cubic\label{sf:long_S_cubic}]{\includegraphics[width=0.3\linewidth]{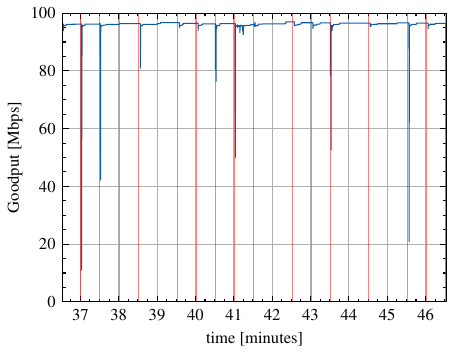}}\hfil
    \subfloat[BBR\label{sf:long_S_bbr}]{\includegraphics[width=0.3\linewidth]{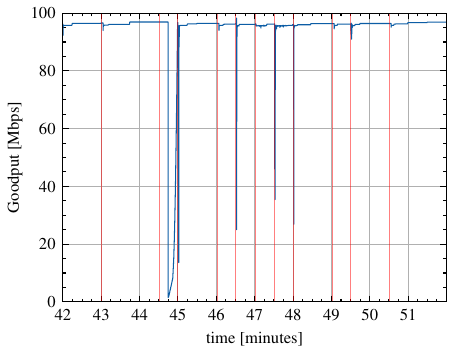}}
    \caption{Evolution of the goodput over 10 minutes of emulation. The time instants at which route changes occur are marked with pink vertical lines.}\vspace{-0.6cm}
    \label{fig:goodput}
\end{figure*}

\begin{figure*}
    \centering
    \subfloat[Reno\label{sf:short_cwnd_reno}]{\includegraphics[width=0.3\linewidth]{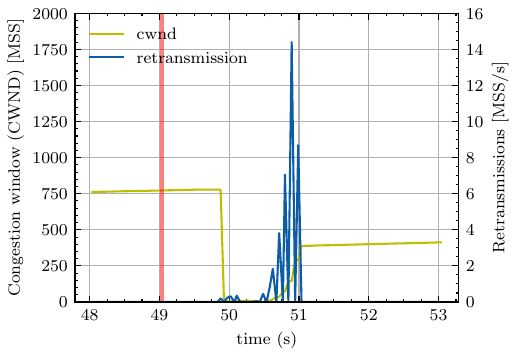}}\hfil
    \subfloat[Cubic\label{sf:short_cwnd_cubic}]{\includegraphics[width=0.3\linewidth]{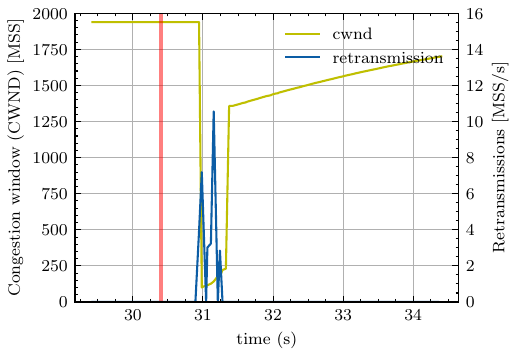}}\hfil
    \subfloat[BBR\label{sf:short_cwnd_bbr}]{\includegraphics[width=0.3\linewidth]{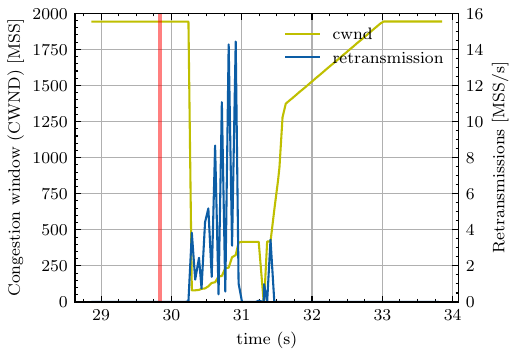}}
    \caption{Evolution of the congestion window and number of retransmissions immediately after a route change, marked with pink vertical lines.}\vspace{-0.6cm}
    \label{fig:short_cwnd}
\end{figure*}

\section{E2E Performance Analysis}
\label{sec:emulator}

In this section, we present our evaluation of the effect of route changes in \gls{leo} satellite networks by using our emulation software under three scenarios:
\begin{enumerate}
\item \textbf{Baseline scenario:} Data packets are transmitted from a server connected to a \gls{gs} in Madrid, Spain, to a client connected to a \gls{gs} in Tokyo, Japan, through the original OneWeb \gls{leo} constellation design, which is a Walker star with $18$ near-polar orbital planes deployed at $1200$\,km above the Earth's surface. An example of the selected route in this scenario is shown in Fig.~\ref{fig:routing}.
    \item \textbf{9 orbital planes:} The same route from Madrid to Tokyo is considered, but the OneWeb constellation is downscaled to $9$ orbital planes instead of the original $18$. This reduces the number of inter-plane \glspl{isl} in the route when compared to the baseline scenario.
    \item \textbf{Aalborg to Cape Town:} Packets are transmitted from Aalborg, Denmark, to Cape Town, South Africa, through the original OneWeb constellation design. This creates routes that mostly consist of intra-plane \glspl{isl}, as the difference in longitude between \glspl{gs} is minimal.
\end{enumerate} 
Each experiment consisted of the continuous transmission during $1000$ seconds between source and destination, where the source possesses a large amount of data to transmit and attempts to use the full capacity of the route. The paths were updated at $30$~s intervals. Such periodic updates have been observed in commercial satellite constellations such as Starlink~\cite{Mohan2023}. The relevant constellation and communication parameters for evaluation are listed in Table~\ref{tab:emulation_parameters}. We note that we did not use Eifel's algorithm or any reordering mitigation technique from the literature, as they are not widely implemented on end-user terminals: the transmitter may not even be aware that they are sending data over a \gls{leo} network, and violating the layer abstraction principle would require pushing updates to billions of devices. Routing is performed at the source \gls{gs} following Dijkstra's algorithm, and the routing metric of choice is the Euclidean distance between satellites. In all cases, we consider that the TCP connection is established between the satellites connecting the source and destination \glspl{gs}.  We observed that, in all three scenarios, the difference between the minimum and maximum propagation time along the established paths is below $2.5$~ms. Furthermore, this jitter affects all packets on the same route equally, and does not cause any reorderings, so TCP can deal with it without any issues. Such a small variation, which is comparable to the behavior of terrestrial mobile networks, confirms that the performance degradation observed in the following is mostly due to the out-of-order delivery due to changes in the paths.

\subsection{Goodput and RTT}

Fig.~\ref{fig:gp_cdf} shows the \gls{cdf} of the goodput  in the baseline scenario. It can be seen that, while \gls{tcp} Reno suffers significantly from the reordering events, Cubic and \gls{bbr} are more robust, operating close to the maximum capacity of the connection. Conversely, from Fig.~\ref{fig:rtt_ccdf}, which shows the complementary \gls{cdf} of the \gls{rtt}, it is seen that the \gls{rtt} with Reno is generally lower than with Cubic and \gls{bbr}, for which the buffers are full most of the time. However, there are significant drops in capacity for all three algorithms: as Fig.~\ref{fig:goodput} shows, even Cubic and \gls{bbr} experience periodic and significant drops right after some route changes, which can have significant effects on applications that require a constant capacity, such as video conferencing or gaming.

We look at these transient effects more in detail by zooming in on the seconds immediately before and after a route change. In Fig.~\ref{fig:short_cwnd} we considered three route changes in which the reordering issue appears. We see that the congestion window tends to grow faster for Cubic and \gls{bbr}, but all three algorithms experience a spike in retransmissions and a significant drop in the congestion window about a second after the route change. Interestingly, Cubic is the fastest to react, recovering after less than half a second, while Reno and \gls{bbr} keep retransmitting packets for more than a second.

\begin{figure*}
    \centering
    \subfloat[Reno\label{sf:short_rtt_reno}]{\includegraphics[width=0.3\linewidth]{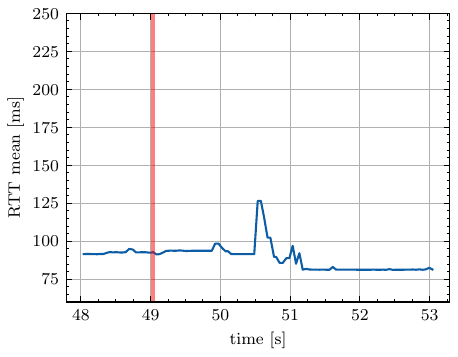}}\hfil
    \subfloat[Cubic\label{sf:short_rtt_cubic}]{\includegraphics[width=0.3\linewidth]{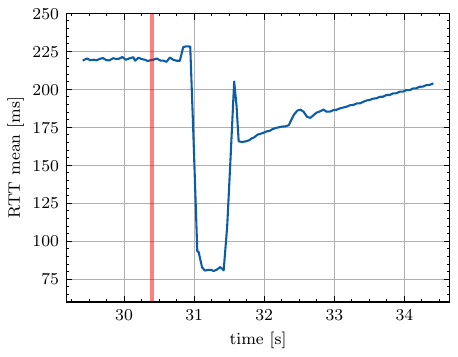}}\hfil
    \subfloat[BBR\label{sf:short_rtt_bbr}]{\includegraphics[width=0.3\linewidth]{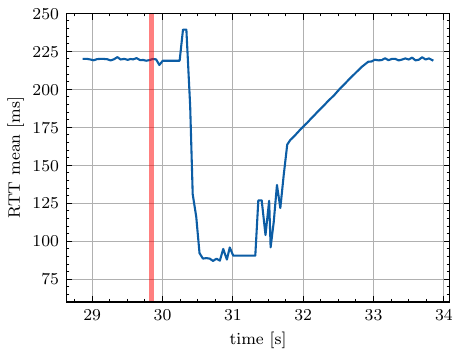}}
    \caption{Evolution of the smoothed \gls{rtt} (blue lines) immediately after a route change, marked with pink vertical lines.}
    \label{fig:short_rtt}
\end{figure*}

Next, we analyze the effect of the route changes on the measured \gls{rtt} at the receiver in the baseline scenario, shown in Fig.~\ref{fig:short_rtt}. As we indicated above, \gls{tcp} Reno has a completely different behavior from the newer algorithms: the \gls{rtt} is very low until the sender enters loss recovery. At that point,  the observed \gls{rtt} increases due to a massive number of retransmissions, and returns to normal (at a slightly lower value) after the sender returns to congestion avoidance mode. On the other hand, Cubic and \gls{bbr} tend to fill the buffer and maintain a higher \gls{rtt}. The measured \gls{rtt} then drops after the route change, when the sender enters the recovery phase, and rapidly increases again. We also note that \gls{bbr} is supposed to avoid creating queues at the bottleneck when not competing with Cubic, but it does so in this scenario, as shown by the increase in the \gls{rtt}. The frequent route changes have the same effect on the max filter driving the congestion control as a fast-varying capacity~\cite{chiariotti2021bbr}, causing \gls{bbr} to overshoot the available capacity and be limited by the congestion window (set at 3 times the measured bandwidth-delay product) rather than by its pacing and increasing the latency correspondingly.

\subsection{Sensitivity analysis}
\begin{figure}[t]
    \centering
    \subfloat[\gls{ccdf} of RTT]{\includegraphics[width=0.7\columnwidth]{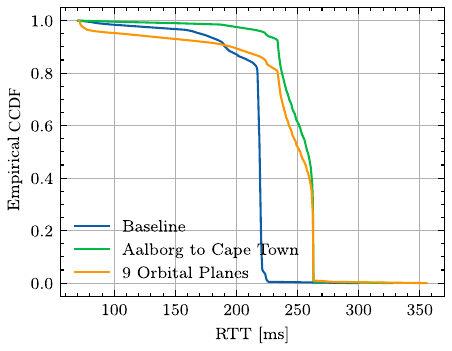}\label{fig:sensitivity_a}}\\
    \subfloat[\gls{cdf} of Goodput]{\includegraphics[width=0.7\columnwidth]{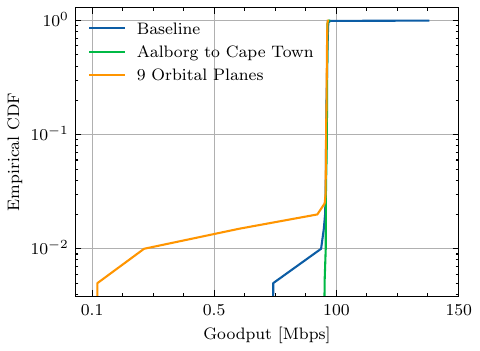}\label{fig:sensitivity_b}}
    \caption{\gls{ccdf} of \gls{rtt} and \gls{cdf} of goodput for the three different scenarios.}\vspace{-0.4cm}
    \label{fig:sensitivity}
\end{figure}

To further investigate the sensitivity of Cubic, the best-performing algorithm in previous tests, to specific route parameters we collected results for the three scenarios, sampling the \gls{rtt} and goodput every 50~ms over multiple simulations for a total observation period of $1000$~s.
Fig.~\ref{fig:sensitivity} shows that the \gls{rtt} of the baseline scenario with \glspl{gs} in Madrid and Tokyo achieves the lowest RTT and the two other scenarios experience a comparable RTT. However, it can also be seen that the RTT in the Aalborg to Cape Town scenario is the most stable. The stability of this route's RTT is reflected in the stability of the goodput, shown in Fig.~\ref{fig:sensitivity_b}, where no goodput variations are observed. On the other hand, it can be seen that the goodput of the baseline scenario and that of the scenario with $9$ orbital planes dropped significantly approximately $0.5$\% and $1$\% of the time, respectively. Thus, even though the average goodput for the three scenarios is comparable, it is evident that even Cubic may experience sporadic issues in some scenarios, causing issues for applications that require a stable high-throughput connection. Thus, the behavior observed in Fig.~\ref{fig:sensitivity_b} highlights the need for mechanisms to improve the stability of the routes utilizing inter-plane ISLs, so these can provide a more stable goodput as those with only inter-plane ISLs.

\section{Conclusions}

Our results show that the reordering problem may be a significant issue for \gls{tcp} in \gls{leo} satellite communications. While the average throughput for long-term flows is almost unaffected for modern congestion control algorithms such as Cubic or \gls{bbr}, the massive reorderings caused by route changes cause significant short-term drops in the sending rate, which may have impacts on the user experience. For example, video services such as \gls{dash}, which operate on top of \gls{tcp}, may experience rebuffering events if the perceived capacity drops sharply due to the sender entering into loss recovery mode unnecessarily, leading to significant user experience issues.

Future work will focus on designing network-level solutions to the reordering problem that exploit the predictable mobility and reconfiguration in the \gls{leo} constellations. Specifically, queue and flow management systems, as well as delayed sending at intermediate satellites, may be used to ``correct'' the reordering before the packets reach the receiver, at the cost of slightly reducing the average throughput and increasing the delay and signaling among satellites. On the other hand, \gls{e2e} solutions may include modifications to the loss detection mechanisms of \gls{tcp}, e.g., shifting from a DUPACK-based loss detection to a timeout-based method: as we mentioned above, this is a complex endeavor due to the massive number of devices that would need to be updated, and in-network solutions would significantly reduce the deployment costs.

\section*{Acknowledgments}
This work was partly supported by the Villum Investigator Grant \mbox{``WATER''} from the Velux Foundation, Denmark, and by the European Union, under the Italian National Recovery and Resilience Plan of NextGenerationEU, as part of the partnership on ``Telecommunications of the Future'' (PE0000001 -- ``RESTART'').
\balance
\bibliographystyle{IEEEtran}
\bibliography{bib}
\end{document}